\documentclass[runningheads]{llncs}
\usepackage[T1]{fontenc}
\usepackage{graphicx}
\usepackage{booktabs} 
\usepackage{array}
\usepackage{tabularx}
\usepackage{multirow}
\usepackage{diagbox} 
\usepackage{lipsum} 
\usepackage[bottom]{footmisc}
\usepackage{hyperref}
\usepackage{appendix}
\usepackage{xcolor}
\usepackage{longtable}

\usepackage[backend=biber,style=numeric,sorting=none]{biblatex} 
\begin{document}
\title{Predicting Learning Performance with Large Language Models: A Study in Adult Literacy}

%
\author{Liang Zhang\inst{1,2} \and
Jionghao Lin\inst{3}
\and 
Conrad Borchers\inst{3} 
\and
John Sabatini \inst{1,4}
\and 
John Hollander\inst{1,4} 
\and
Meng Cao \inst{4}
\and 
Xiangen Hu\inst{4,5}}
\authorrunning{L. Zhang et al.}
%
\institute{Institute for Intelligent Systems, University of Memphis, Memphis, TN 38152, USA \and
Department of Electrical and Computer Engineering, University of Memphis, Memphis, TN 38152, USA \\
\email{lzhang13@memphis.edu}
\\
\and Human-Computer Interaction Institute, Carnegie Mellon University, Pittsburgh, PA 15213, USA\\
\email{\{jionghao,cborcher\}@cs.cmu.edu}
\\
\and
Department of Psychology, University of Memphis, Memphis, TN 38152, USA \\
\email{\{jpsbtini,jmhllndr,mcao\}@memphis.edu}\\
\and Department of Applied Social Sciences, Hong Kong Polytechnic University, Hong Kong, PR China\\
\email{xiangen.hu@polyu.edu.hk}
\\
}
\maketitle              

\begin{abstract} 

Intelligent Tutoring Systems (ITSs) have significantly enhanced adult literacy training, a key factor for societal participation, employment opportunities, and lifelong learning. Our study investigates the application of advanced AI models, including Large Language Models (LLMs) like GPT-4, for predicting learning performance in adult literacy programs in ITSs. This research is motivated by the potential of LLMs to predict learning performance based on its inherent reasoning and computational capabilities. By using reading comprehension datasets from the ITS, AutoTutor, we evaluate the predictive capabilities of GPT-4 versus traditional machine learning methods in predicting learning performance through five-fold cross-validation techniques. Our findings show that the GPT-4 presents the competitive predictive abilities with traditional machine learning methods such as Bayesian Knowledge Tracing, Performance Factor Analysis, Sparse Factor Analysis Lite (SPARFA-Lite), tensor factorization and eXtreme Gradient Boosting (XGBoost). While XGBoost (trained on local machine) outperforms GPT-4 in predictive accuracy, GPT-4-selected XGBoost and its subsequent tuning on the GPT-4 platform demonstrates superior performance compared to local machine execution. Moreover, our investigation into hyper-parameter tuning by GPT-4 versus grid-search suggests comparable performance, albeit with less stability in the automated approach, using XGBoost as the case study. Our study contributes to the field by highlighting the potential of integrating LLMs with traditional machine learning models to enhance predictive accuracy and personalize adult literacy education, setting a foundation for future research in applying LLMs within ITSs. 

\end{abstract} 
\section{Introduction}

Adult literacy education, particularly in reading comprehension, empowers individuals to fully participate in society, access better job opportunities, and engage in lifelong learning \cite{national2012improving,barton2012literacy}. Effective literacy education training programs are designed to address the diverse needs of adult learners, incorporating strategies that enhance reading skills, comprehension, and the ability to critically analyze texts \cite{woolley2011reading,shi2018diagnostic}. Notably, the effectiveness of these programs often relies on the accurate assessment and continuous improvement of personalized instructions to meet learner needs \cite{shi2018diagnostic,fang2019conversation}. In this context, the prediction of learning performance is important, which can allow for the early identification of individuals who may require additional support, enabling interventions that are precisely targeted to enhance reading comprehension and literacy skills. 

ITSs are one common form of personalizing instruction in adult literacy. ITSs are computer-based systems that tracks and assesses learning progress and further facilitates the adaptation of instruction to better meet learners' needs \cite{graesser2004autotutor,nye2014autotutor,graesser2021using}. A significant component of one ITS is predicting learning performance based on machine learning models to personalize instruction \cite{corbett1997intelligent,graesser2012intelligent}. Learning performance prediction relies on analyzing and modeling historical data, including records of learners' correctness in problem-solving attempts \cite{corbett1994knowledge,pavlik2021logistic}. Advanced machine learning models using these fine-grain data, alongside natural language related to the problem-solving context in adult literacy education, could potentially improve learner performance prediction and improve the personalization of instruction. However, to unlock the potential of ITS in adult literacy education, advanced AI models, such as multimodal machine learning models and LLMs, to accurately predict the learning performance are under-explored. 

Recent advancements in AI models, such as LLMs, have demonstrated remarkable predictive capabilities, including mathematical reasoning \cite{imani2023mathprompter,ahn2024large} and time series forecasting \cite{jin2023time,gruver2023large,zhang2024large}. These achievements underscore the LLMs' potentials in understanding patterns in data relevant to learner modeling tasks. In the domain of education, prior research also demonstrated the potentials of leveraging LLMs for predictive analysis (e.g., predicting learning performance in computer science education \cite{liu2022open} and identifying at-risk learners \cite{susnjak2023beyond}). Despite these advances \cite{liu2022open, susnjak2023beyond}, the application of LLMs in enhancing predictive analytics within ITSs remains in its early stages. Motivated by the proven effectiveness of LLMs in educational prediction tasks, our study explores the potential of LLMs, specifically GPT-4 \cite{achiam2023gpt}, in comparison with traditional methods such as Bayesian Knowledge Tracing (BKT) \cite{corbett1994knowledge,yudelson2013individualized}, Performance Factor Analysis (PFA) \cite{pavlik2009performance}, Sparse Factor Analysis Lite (SPARFA-Lite) \cite{lan2014quantized}, tensor factorization \cite{doan2019rank} and eXtreme Gradient Boosting (XGBoost) \cite{chen2015xgboost,asselman2023enhancing}, for predicting learner performance in the context of adult literacy education. The present study investigates two \textbf{R}esearch \textbf{Q}uestions:
\begin{itemize}
    \item[] \textbf{RQ1:} How effectively can the GPT-4 model, through specific prompting strategies, predict learning performance in adult literacy programs compared to existing benchmark models?
    \item[] \textbf{RQ2:} How can GPT-4 augment traditional human-led efforts in enhancing the prediction accuracy of learning performance in adult literacy lessons (e.g., \textit{Persuasive Text}, \textit{Cause and Effect},\textit{Problems and Solution})?
\end{itemize}

\begin{figure}[htbp]
\centering
\includegraphics[width=0.75\textwidth]{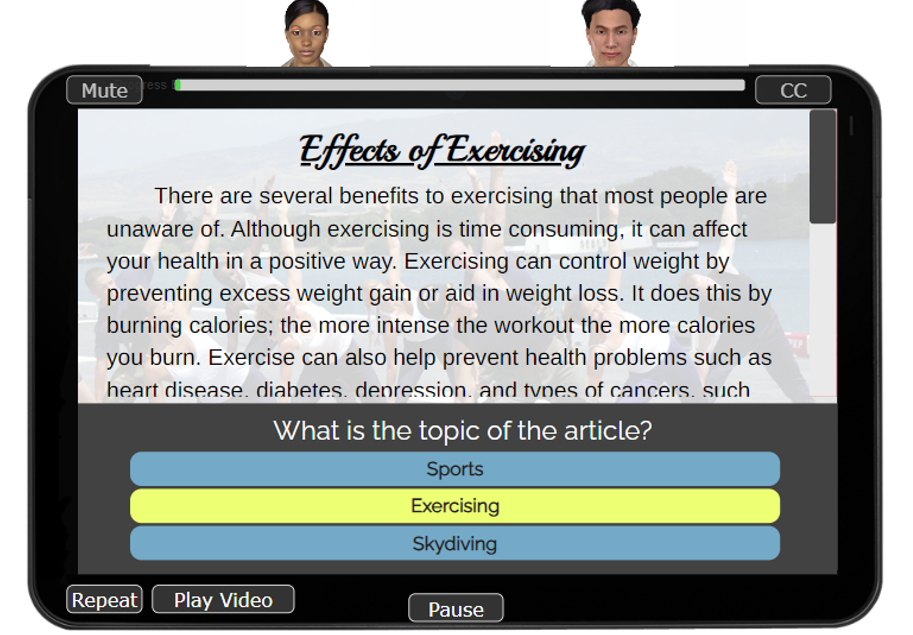}
\caption{The interface of AutoTutor for the Center for the Study of Adult Literacy.} \label{fig:auto_tutor_csal}
\end{figure}

Our study utilized the reading comprehension datasets collected from the well-known ITS, AutoTutor, developed for the Center for the Study of Adult Literacy (CSAL) \cite{graesser2016reading}. Fig. \ref{fig:auto_tutor_csal} illustrates the interface of the lessons developed using CSAL AutoTutor. The datasets comprise attributes such as learner ID, questions, attempts, and learners' performance scores for each lesson. To answer \textbf{RQ1}, we employed the widely-used models, including BKT, PFA, SPARFA-Lite and XGBoost, comparing them with the GPT-4 model \cite{achiam2023gpt}. We assessed model performance through five-fold cross-validation for all models, revealing that the XGBoost model surpassed GPT-4 in predicting learning performance. Interestingly, when prompted GPT-4 for model selection, GPT-4 itself recommended XGBoost for predicting learning performance. Running the GPT-4-selected XGBoost model on the GPT-4 platform yielded superior results compared to its execution on a local machine. We then answered \textbf{RQ2} by examining the tuning of hyper-parameters by GPT-4 versus manual tuning. Specifically, we prompted GPT-4 to optimize an XGBoost model for predicting learning performance, focusing on enhancing its prediction capabilities. In parallel, we manually adjusted the XGBoost model's hyper-parameters, such as the number of trees, learning rate, and maximum tree depth, through a grid-search method on a local machine for comparison. Our findings indicated that while the GPT-4-tuned hyper-parameters achieved performance comparable to that of manually tuned models, they exhibited less stability than those optimized through manual grid search. 

\section{Related Work}
\subsection{Adult Literacy Education in Intelligent Tutoring Systems} \label{sub:its}

Adult literacy education has long advocated the computer aided adaptive learning technologies for their capacity to provide personalized and cost-effective educational solutions \cite{graesser2019educational}. These technologies are commonly applied in ITSs, as previously mentioned. ITSs can offer personalized tutoring and adaptive instructions tailed to the individual learner, dynamically adjusting the difficulty levels and contents of lessons based on the learner's responses to questions and tasks \cite{graesser2021using,graesser2017two,fang2022patterns}. For instance, utilizing web-based applications, ITSs deploy computer-based agents to deliver customized reading materials and learning tasks, ensuring support is readily available for learners facing challenges \cite{graesser2016reading,graesser2019educational,fang2018clustering}. The significant impact of ITSs lies in their ability to create an adaptive learning environment that supports and responds to individual educational needs, thereby empowering learners to advance at their own pace. Furthermore, ITSs tackle the issue of scarce human tutoring and classroom resources for adult learners by utilizing these systems to improve reading comprehension skills \cite{graesser2019educational,lippert2019using,rose2019technology}. 

A notable ITS prototype is the previously mentioned CSAL AutoTutor, specifically developed for enhancing reading comprehension. The CSAL AutoTutor employs the trialogue design, which includes one human learner and two computer agents (virtual tutor and virtual companion) (see Fig. \ref{fig:auto_tutor_csal}) \cite{graesser2016reading}. The interaction between computer agents and learners including chat and talking heads \cite{graesser2017assessment}. These agents guide learners towards their learning goals via conversation. The system assesses learners' responses, provides feedback, matches expectation and corrects misconceptions, which is considered the Expectation-Misconception Tailing (EMT) principle \cite{graesser2005computerized}. Once all lesson expectations are met, the tutoring session concludes. 

Generally, AutoTutor has been shown to significantly enhance reading effectiveness, with studies indicating an average learning gain of 0.8 standard deviations over traditional teaching methods \cite{nye2014autotutor}. Fang et al. observed that AutorTutor markedly benefits individuals with low literacy levels \cite{fang2019conversation}. Shi et al. identified AutoTutor as an effective and comprehensive tool for assessing and supporting the improvement of adult literacy skills \cite{shi2018exploring}. Additionally, research into individual learning differences in reading comprehension within ITS environments has shown that learning performance adheres closely to power-law functions, demonstrating positive learning rates facilitated by AutoTutor \cite{zhang2023exploring}. 

\subsection{Learning Performance Prediction}

Learning performance prediction is an important task in the field of ITS in education. By understanding the learner's performance, ITS can accurately assess learning states and offer tailored instructions to support learners throughout learning process, particularly when they encounter difficulties with questions, face early risks of failure, or experience wheel-spinning \cite{beck2013wheel,gong2014student}. 

The predictive task for learning performance utilizes historical records to predict future performance on questions, incorporating data from multiple attempts \cite{desmarais2012review,pavlik2013tutoring,pavlik2023automated}. Driven by the needs of high accurate model for learning performance prediction, many previous works employed machine learning methods including BKT, PFA, SPARFA-Lite and tensor factorization. Widely recognized predictive models such as BKT \cite{corbett1994knowledge,yudelson2013individualized} and PFA \cite{pavlik2009performance,pavlik2021logistic}, leveraging Bayesian networks and logistic regression for learner performance prediction respectively. BKT outlines four probabilistic parameters: ``known'' (initial or prior knowledge), ``slip'' (incorrectly answering despite knowing the skill), ``guess'' (correctly answering without knowing the skill), and ``learn'' (mastering a skill in subsequent practices) \cite{yudelson2013individualized}. PFA, on the other hand, includes parameters that account for prior success and failures in answering questions, skill difficulty reflecting the inherent challenge of the skill, and individual learning rates indicating how fast the learner improve in mastering knowledge \cite{pavlik2009performance,chi2011instructional}. Both methods have been utilized for predicting learning performance owing to their stability, strong predictive performance, and explainability \cite{yudelson2013individualized,pelanek2017bayesian,pavlik2023automated}. SPARFA-Lite utilizes quantized matrix completion to predict learner performance in knowledge tracing, representing the probability of answering questions successfully based on three factors: 1) the learner's understanding of latent concepts, 2) the relationship between questions and concepts, and 3) the inherent difficulty of each question \cite{lan2013sparse}. The tensor factorization method structurally represents learner knowledge in a three-dimensional space, incorporating critical factors such as learners, questions, and attempts to influence learning progress. This approach calculates probability estimates for learner performance using mathematical tensor factorization. 

\subsection{Large Language Models in Education}

LLMs, pre-trained on massive amounts of data, enabling them to generate human-like text, answer questions, and perform reasoning tasks with unprecedented accuracy \cite{huang2022towards,wei2022chain}. LLMs like ChatGPT have demonstrated remarkable advancements in AI, driving revolutionary shifts in education applications through enhancing instructional feedback \cite{dai2023can,schmucker2023ruffle,lin2024improving}, boosting student engagement \cite{tan2023large}, and offering personalized learning experiences \cite{xiao2023evaluating}. 

However, the applications of LLMs in enhancing predictive analytics within ITSs remains in its early stages. Liu et al.'s investigation \cite{liu2023evaluating} on ChaGPT's effectiveness in logical reasoning, particularly in making prediction-based inferences for multiple-choice reading comprehension and natural language inference tasks, highlights its adeptness at complex educational reasoning challenges. Liu et al. \cite{liu2022open} has incorporated the ChatGPT for open-ended knowledge tracing in computer science education, enabling enhanced prediction of code snippets for open-ended response analysis. Susnjak \cite{susnjak2023beyond} has attempted to integrate ChatGPT with machine learning models, enabling advanced predictive analytics to assist at-risk learners through evidence-based remedial recommendations. Further instances will not be elaborated upon. These cases highlight the advanced predictive capabilities of LLMs or their collaboration with machine learning models for predictive tasks, inspiring further exploration of LLMs' potential in advancing educational predictive applications. 

\section{Methods}

\subsection{Dataset}

In this study, we utilized datasets from AutoTutor lessons developed for the Center for the Study of Adult Literacy (CSAL), which is public accessible online\footnotemark[1]\footnotetext[1]{AutoTutor Moodel Website: \href{https://sites.autotutor.org/}{https://sites.autotutor.org/}; Adult Literacy and Adult Education Website: \href{https://adulted.autotutor.org/}{ https://adulted.autotutor.org/}}. Our study was granted ethical approval with the Institutional Review Board (IRB) number: H15257. As described in subsection \ref{sub:its}, the CSAL AutoTutor employs a trialogue interaction mode involving two computer agents, a tutor agent and a virtual peer agent, to facilitate human learners' acquisition of reading comprehension skills through multiple-choice questions \cite{graesser2016reading}. The selected lessons for our analysis include ``\textit{Persuasive Text}'' (Lesson 1), ``\textit{Cause and Effect}'' (Lesson 2), ``\textit{Problems and Solution}'' (Lesson 3). Table \ref{tab:all_dataset} presents the basic statistics about the dataset on learner performance, detailing information about the learners, questions, and attempts for each lesson. 

\begin{table}[ht] 
\centering
\renewcommand{\arraystretch}{1}  
\caption{Dataset from the CSAL AutoTutor lessons}
\label{tab:all_dataset} 
\begin{tabularx}{\textwidth}{>{\hsize=0.6\hsize}X>{\hsize=1.2\hsize}X>{\hsize=1.1\hsize}c>{\hsize=1.1\hsize}c>{\hsize=1.1\hsize}c} 
\toprule
Dataset & Lesson Name & \# Learners & \# Questions & Max. Attempt \\ 
\midrule
Lesson 1 & Persuasive Text & 66 & 8 & 9 \\
Lesson 2 & Cause and Effect & 68 & 9 & 9 \\
Lesson 3 & Problems and Solution & 86 & 11 & 5 \\ 
\bottomrule 
\end{tabularx}
\end{table}

\subsection{The Proposed LLM-based Prediction Method}

\begin{figure}[htbp]
\includegraphics[width=\textwidth]{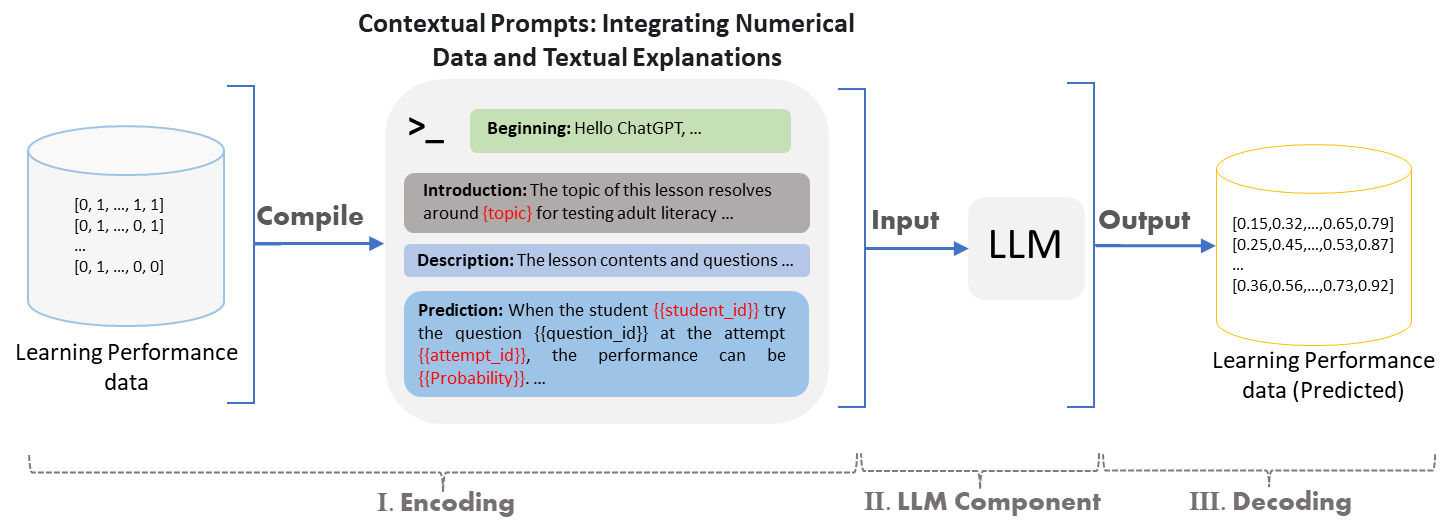}
\caption{LLM-based prediction framework for learner learning performance.} \label{fig:LLM_prompt}
\end{figure}

We developed a LLM-based framework to trace and predict learner performance, as illustrated in Fig. \ref{fig:LLM_prompt}. This framework includes three procedures: 1) encoding for converting numerical value to contextual prompts, 2) the LLM component for analyzing these prompts and executing predictions, and 3) decoding for outputting the prediction information and assessment along with interpretations. 

\textbf{Learning Performance Data.} learner performance on question-answering tasks was recorded as binary data (labeled as correct or incorrect), to reflect the learner's success or failure in answering the questions. The records also captured the number of attempts made by the learner. For instance, the performance of the learner on the \(i^{th}\) question during their \(j^{th}\) attempt is recorded as 1 for a correct answer and 0 for an incorrect one (this setting applies to the training dataset, whereas, for the testing dataset, performance data are omitted to enable future prediction). Our objective is to input these performance data into LLM to identify data patterns and latent learning features such as initial knowledge level and learning rate, aiming to predict the learner's future performance based on the learner's historical attempts. 

\textbf{Encoding.} Learning performance in binary indicator variables representing correct or incorrect responses can be compiled into a contextual prompt by integrating numerical data with textual explanations. For example, the entry \(x_{l,i,j}\) could be represented as ``The current learner \(l\) attempted to answer the \(i^{th}\) question titled as `...' on their \(j^{th}\) attempt. Their performance was observed as 1 or 0''. Essential considerations for this encoding process include: \textbf{(a)} \textit{Numerical to Text Conversion}: This involves the embedding of numerical value along with the question contents and answers into a narrative or textual format. \textbf{(b)} \textit{Contextual Integration:} This aspect involves incorporating information related to the lesson topic, content, and questions, along with knowledge gained from reading comprehension, to enrich understanding of both question and lesson material. This enrichment process also facilitates tailoring and assessing reading comprehension skills of the learner in the learning process. 

\textbf{LLM Component.} The contextual prompt serves as input into the LLM component for data analysis and modeling, aimed at predicting learner performance on new or repeated question attempts. Two primary prediction strategies are encompassed in this component: 1) leveraging the inherent reasoning, understanding, and computational capabilities of the GPT-4 model, e.g., the heuristic-based approach; and 2) utilizing available machine learning models, which are automatically selected and fine-tuned by the GPT-4 model for predictive tasks. Through extensive analysis of trial experiments, the following assumptions are included: 
\textbf{(a)} GPT-4 can be pre-trained in predictive task, \textbf{(b)} GPT-4 possesses the capability to extract domain-specific knowledge from distinct questions and attempts, \textbf{(c)} GPT-4 can uncover latent learning features within contextual performance data, enabling the evaluation of learners' reading comprehension skills, and \textbf{(d)} Information inferred by GPT-4, alongside other computational models, can be leveraged to discern trends, patterns, and predict learner learning performance. 

\textbf{Decoding.} In the decoding phase, predictive outcomes regarding learning performance are produced through a mechanism that employs either heuristic approaches or machine learning models, which GPT-4 automatically selects and implements. These predictions draw on an analysis of the learning performance distribution integrated in previously mentioned contextual prompts. For instance, it reformat output information into a structured format, such as \textit{``\{`learner ID':..., `Question ID':..., `Attempt':..., `Prediction':..., `Assessment':...\}''}, efficiently conveying the prediction details. This procedure incrementally increases the dimensionality of the data until it aligns with the original input size of the test dataset. 

\subsection{Baseline Methods}

This study will employ machine learning models specifically for predicting learning performance in education \cite{abdelrahman2023knowledge}, as baselines. Below is a concise overview of each selected method. 

\textbf{Bayesian Knowledge Tracing (BKT).} BKT is a computational model used to track and predict learners' knowledge acquisition over time within educational software, particularly within ITSs \cite{yudelson2013individualized,pardos2010modeling,pardos2011kt}. Fundamentally, BKT is based on the principles of Bayesian probability which estimates the likelihood that a learner has mastered a particular skill or concept at various points throughout the learning process, adjusting these estimates in response to the learner's performance on related tasks or questions \cite{pardos2010modeling}.
According to \cite{corbett1994knowledge}, the classic BKT's key parameters include the initial probability of mastering the skill, the probability of acquiring knowledge the learner by transforming from the unmastered state on one skill to master state, the probability of making incorrect answer response by slipping in mastered state on a skill, and the probability of making correct answer response by guessing in unmastered state on a skill. BKT advanced this framework by incorporating customized parameters for each learner and each skill into its modeling and predictions \cite{pardos2010modeling}. By updating these parameters using Bayesian inferences based on each learner's responses to questions over time, BKT refines its estimates of a learner’s knowledge state at a particular time step as the learner responds to questions \cite{corbett1994knowledge,pardos2010modeling}. 

\textbf{Performance Factor Analysis (PFA).} The PFA utilizes logistic regression to predict the learner's performance on the questions by incorporating factors on individual learning ability, skill-related features (e.g., difficulty), and the learner's previous success and failures \cite{pavlik2021logistic,pavlik2009performance,chi2011instructional,yudelson2011user,eglington2022optimize}. Many studies have established PFA as a competitive approach in predicting learner performance, acknowledging the importance of individual differences across skills and learners \cite{pavlik2021logistic,pavlik2023automated,gong2010comparing}.  In our research, we have further refined the PFA model to better account for the variability among individual learners. 

\textbf{Sparse Factor Analysis Lite (SPARFA-Lite).} The SPARFA-Lite, a variant of the Sparse Factor Analysis (SPARFA), employs matrix completion technique to analyze quantified, graded learner performance on questions \cite{lan2014quantized}. This model offers improved efficiency in automatically exploring the number of Knowledge Components for predicting learner performance compared with the traditional compared to the traditional Sparse Factor Analysis model \cite{lan2014quantized}. 

\textbf{Tensor Factorization:} The tensor factorization method decomposed a three-dimensional tensor (representing dimensions of learners, questions and attempts) into a factor matrix for learners and latent features, and a factor tensor that encompasses dimensions of latent features, questions, and attempts \cite{doan2019rank,wang2021knowledge,zhang2023exploring,zhang20243dg}. The latent feature dimensions obtained through tensor decomposition capture learner-specific characteristics such as learning abilities and personalities \cite{wang2021knowledge}. The factor tensor models the knowledge space related to learner interactions with questions. Our prior studies have demonstrated its significant efficacy in predicting learner performance, particularly within the adult literacy domain \cite{zhang2023exploring,zhang20243dg}. Here, the rank-based constraint was applied to regulate the factorization computing \cite{doan2019rank}.

\textbf{eXtreme Gradient Boosting (XGBoost):} The XGBoost algorithm has become an effective model for knowledge tracing, significantly enhancing prediction performance \cite{asselman2023enhancing,su2023xgboost}. At its core, XGBoost constructs an ensemble of decision trees in a sequential manner \cite{chen2015xgboost,chen2016xgboost}. In this process, each subsequent tree is specifically trained to address and correct the residuals or errors made by the preceding tree, effectively enhancing the model's predictive accuracy over iterations. Guided by a gradient descent algorithm, XGBoost optimizes a predefined loss function, systematically reducing prediction errors. Its capability to interpret input features, such as unique learners, questions, and attempts in our study, enables an in-depth understanding of model predictions, enhancing transparency and trustworthiness in predictive analytics. 

\subsection{Evaluation}

We employed the recognized quantitative metric Root Mean Square Error (RMSE), which aligned with peer studies \cite{yudelson2013individualized,pavlik2021logistic,pavlik2023automated,xiong2016going,gervet2020deep}. RMSE provides a measure of the square root of the average squared differences between predicted and actual values \cite{hyndman2006another}. Additionally, we conducted a five-fold cross-validation to obtain RMSE values for comparative analysis. In our LLM-based prediction method, specifically utilizing GPT-4, we allocated four out of five folds for training to enable the model to learn from historical data through contextual prompts. The remaining fold was used for testing, to evaluate the accuracy of predictions made by the trained GPT-4. 

\section{Results}

\subsection{Results on RQ1} 

We investigated the comparison of prediction performance between LLM-based models and Baseline Methods. The result is shown in Table \ref{tab:rmse_metric} which presents the RMSE values of all models across three CSAL lessons, with lower value indicating better performance in prediction. It should be noted that all the RMSE scores were obtained from models after adjusting their hyper-parameters. 

Specifically, the XGBoost (selected by GPT-4) showcases an enhanced application, leveraging GPT-4's strengths in reasoning, computation, and automatic tuning to refine outcomes. Aware of the potential for errors or instability in GPT-4's predictions, we established the reliability of our RMSE by conducting seven repeated prediction runs for both the standard GPT-4 and the GPT-4 enhanced with XGBoost, across each lesson. The outstanding predictive accuracy of the XGBoost (selected by GPT-4) model led us to perform an extensive manual grid search, documented in Table \ref{tab:rmse_metric}. This process entailed evaluating 1,296 combinations of hyper-parameters, including number of trees, learning rate, maximum tree depth, training instance subsample ratio, column subsample ratio per tree, minimum loss reduction for further partitioning, and the minimum sum of instance weight required in a child node. This rigorous hyper-parameter tuning was aimed at further enhancing the model’s performance for each lesson, as detailed in Table \ref{tab:hyperparameter_tuning}.

Table \ref{tab:rmse_metric} presents the RMSE values of all models across three CSAL lessons, with lower value indicating better performance in prediction. The RMSE value of GPT-4, as shown in Table \ref{tab:rmse_metric}, surpasses that of most other models, demonstrating only higher values when compared to XGBoost in Lessons 1 and Lesson 2. Notably, the RMSE value of XGBoost (selected by GPT-4) is the lowest among all six models across the three lessons. This demonstrates the substantial enhancement in predicting learning performance achieved through the integration of GPT-4 and XGBoost. Additionally, the XGBoost model outperforms other traditional knowledge tracing models, including Individualized BKT, PFA, SPARFALite, and Tensor Factorization, in the reading comprehension data. As for the standard errors for the RMSE values, lower values indicate less variability in the estimated RMSE values, thereby suggesting greater confidence in the accuracy of the predictions. The standard errors for all RMSE values related to GPT-4 and XGBoost (as chosen by GPT-4) fall within the range of [0.004, 0.009], indicating a relatively moderate variability in the prediction outcomes across all models. 

\begin{table}[ht]
\caption{Comparison of model performance using RMSE with the standard error from five-folds cross validation}
\label{tab:rmse_metric}
\begin{tabularx}{\textwidth}{@{}l*{3}{X}@{}}
\toprule 
\diagbox{Models}{Lessons} & \begin{tabular}{@{}l@{}}Lesson 1 \\ (RMSE)\end{tabular}  & \begin{tabular}{@{}l@{}}Lesson 2 \\ (RMSE)\end{tabular}  & \begin{tabular}{@{}l@{}}Lesson 3 \\ (RMSE)\end{tabular}  \\
\midrule
BKT & \(0.430_{0.004}\) & \(0.375_{0.009}\) & \(0.392_{0.006}\) \\
PFA & \(0.440_{0.015}\) & \(0.408_{0.005}\) & \(0.407_{0.012}\) \\
SPARFA-Lite & \(0.603_{0.039}\) & \(0.522_{0.017}\) & \(0.460_{0.015}\) \\
Tensor Factorization & \(0.437_{0.011}\) & \(0.385_{0.009}\) & \(0.395_{0.011}\) \\
XGBoost & \(0.412_{0.010}\) & \(0.366_{0.005}\) & \(0.384_{0.011}\) \\
GPT-4 & \(0.415_{0.004}\) & \(0.370_{0.007}\) & \(0.381_{0.009}\) \\
XGBoost (selected by GPT-4) & \textbf{\(0.398_{0.008}\)} & \textbf{\(0.351_{0.006}\)} & \textbf{\(0.381_{0.008}\)} \\
\bottomrule
\end{tabularx}
\end{table}

\subsection{Results on RQ2}

\begin{table}[ht]
\centering
\caption{Comparison of RMSE  of hyper-parameter tuning methods for XGBoost (Note: the Std. is the abbreviation of standard deviation).}
\label{tab:hyperparameter_tuning}
\setlength{\extrarowheight}{2pt} 
\begin{tabularx}{\textwidth}{p{3.7cm} *{5}{>{\centering\arraybackslash}p{1.38cm}} c}
\toprule
\textbf{Methods} & \textbf{Lessons} & \textbf{Mean} & \textbf{Median} & \textbf{Std.} & \textbf{Min.} & \textbf{Max.} \\ \midrule
\multirow{3}{3cm}{\centering \begin{tabular}{@{}l@{}}Hyper-parameters tuning \\ by GPT-4\end{tabular}} 
& Lesson 1 &  0.435 &  0.422 & 0.053 & 0.398 & 0.552 \\
& Lesson 2 &  0.376 &  0.361 & 0.033 & 0.351 & 0.444 \\
& Lesson 3 &  0.398 &  0.382 & 0.036 & 0.381 & 0.480 \\ \midrule
\multirow{3}{3cm}{\centering \begin{tabular}{@{}l@{}}Hyper-parameters tuning \\ by manual grid search \end{tabular}}
& Lesson 1 & 0.433 & 0.426 & 0.017 & 0.412 & 0.484 \\
& Lesson 2 & 0.391 & 0.384 & 0.020 & 0.366 & 0.423  \\
& Lesson 3 & 0.396 & 0.394 & 0.010  & 0.384 & 0.433 \\ \bottomrule
\end{tabularx}
\end{table}

Table \ref{tab:hyperparameter_tuning} displays a comparative analysis of RMSE values from two hyperparameter tuning approaches for the XGBoost model: one selected by GPT-4 and the other via manual grid search. GPT-4 consistently yields lower minimum and median RMSE values across all three lessons compared to the manual method. However, GPT-4's method results in a lower mean RMSE value only for Lesson 2. The standard deviation values from GPT-4 are larger than those from the manual grid search, indicating a wider variability in RMSE outcomes. Additionally, the minimum and maximum range of values obtained through GPT-4's method exceed those from the manual approach, suggesting a greater spread in the performance results.

\section{Discussions}

\subsection{Efficient LLM-based Method for Predicting Learning Performance}

Our study highlights the capabilities of GPT-4 in predicting learning performance in ITSs. We provide an in-depth examination of our experimental results regarding the application of GPT-4 for predicting learning performance within the CSAL AutoTutor datasets. Two exact predictive strategies, one leveraging the inherent heuristic reasoning method and the other utilizing available machine learning models, are implemented by GPT-4. 

When employing its heuristic-based reasoning approach, GPT-4 takes into account factors such as the perceived difficulty of questions and their attempt frequency. This approach does not rely on a fixed algorithm but uses logical reasoning to analyze historical performance data. It assumes that questions deemed more difficult are less likely to be answered correctly on the first try. Furthermore, if learners make multiple attempts on certain questions, GPT-4 interprets this as a sign of struggle with the material, leading to a more conservative performance prediction. 

At the same time, enhancing GPT-4's predictive accuracy significantly involves incorporating reading comprehension materials, questions, and additional background information to craft context-specific prompts. By understanding the learning content and the questions' context, GPT-4 can offer explanations and leverage its vast knowledge base more effectively. This contextually enriched reasoning allows GPT-4 to outperform traditional learning performance prediction methods, such as BKT, PFA, SPARFA-Lite, and Tensor Factorization. The result is not just more accurate predictions, but also insights that are directly relevant and tailored to the specific learning scenario. This makes GPT-4 an invaluable tool for educators seeking to understand and improve student learning performance. 

When utilizing available machine learning models, GPT-4 demonstrates its ability by recommending and applying a range of machine learning models tailored to the specific needs of the data. Among these models are logistic regression, random forest, gradient boosting machine, and XGBoost. GPT-4's unique self-programming ability enables it to autonomously test these models and select the most effective one based on performance metrics from validation results. Through this process, XGBoost is identified as the most suitable model for predicting learning outcomes, leading to a novel approach in our experiments that combines the strengths of GPT-4 with XGBoost, referred to as GPT-4 with selected XGBoost. This approach remains adaptable, with GPT-4 continuously seeking to refine and enhance its choice of models. The fusion of GPT-4's capabilities with advanced machine learning techniques broadens its application scope, pushing the boundaries of what can be achieved in computational tasks. This not only showcases GPT-4's potential for complex problem-solving but also highlights its role in driving forward the evolution of ITSs.

\subsection{Prompt Strategy for Predicting Learning Performance}

In this study, the prompt engineering plays a crucial role. The foundational framework of our prompts encompasses encoding for the contextual representation of numerical values and decoding to facilitate LLM-based understanding, reasoning, and analysis in the generation of predictive outputs. This approach allows the LLM to seamlessly integrate all processes, from data input to final prediction. It employs self-search and self-optimization for refining prompt engineering, alongside semantic compiling techniques for processing learning performance data. The Chain-of-Thought prompt strategy \cite{wei2022chain} is employed to generate GPT-4 output that illustrates model reasoning and its interpretative process. By activating specific prompts within GPT-4, we guide it to more effectively analyze and interpret learner learning performance data. This method not only improves the transparency of the AI's decision-making process but also enhances the precision and relevance of its predictive capabilities. 

Specifically, the Chain-of-Thought prompt strategy systematically maps out the reasoning steps necessary for predicting learning performance, employing a sequence of precisely tailored prompt compositions to ensure effective execution of each steps. These compositions encompass: \textbf{(a)} \textit{Presentation of Learning Materials:} Share the learning materials and associated comprehension questions to establish a basis for analysis. \textbf{(b)} \textit{Contextual Transcriptions of Learning Performance Data:} Provide a detailed contextual representation of the learning performance data. \textbf{(c)} \textit{Analysis Request:} Clearly articulate the request for data analysis, specifying the desired insights or outcomes. \textbf{(d)} \textit{Method Selection:} GPT-4 suggests appropriate analytical or machine learning methods based on the project needs. \textbf{(e)} \textit{Model Development:} Assistance in developing a machine model, e.g., XGBoost, including training and validation across dataset folds. \textbf{(f)} \textit{Performance Evaluation:} Calculation and presentation of validation outcomes, such as RMSE, for each fold. \textbf{(g)} \textit{Configuration Disclosure:} Detailed sharing of the model's configuration settings for transparency and reproducibility. \textbf{(h)} \textit{Skill Assessment:} Discussion on assessing learners' reading comprehension skills based on their performance data. \textbf{(i)}\textit{ Optimization:} Guidance on fine-tuning the model's hyperparameters for improved predictive performance. \textbf{(j)} \textit{Iterative Feedback:} Continuous exchange for clarification, refinement, and further analysis based on user inputs and GPT-4's suggestions. For a comprehensive overview and detailed instructions, please refer to \hyperref[appendix:A]{Appendix A}.

\section{Limitations}

Although the present study highlights the potential of LLMs in enhancing predictive accuracy of learning performance prediction, it also identifies certain limitations. Future work remains in strengthening the connections between specific reading comprehension knowledge and skills and the reasoning process. There is a need to explore how these connections can be utilized to refine prompts and enhance predictive effectiveness. Specifically, constraints related to the fine-tuning of LLM-based platforms or APIs may hinder the optimization of models tailed for our dataset. Additionally, limitations in executing deep learning models restrict the application of advanced techniques such as Deep Knowledge Tracing (DKT) \cite{piech2015deep}, Self-Attentive Knowledge Tracing (SAKT) \cite{pandey2019self}, Dynamic Key-Value Memory Networks (DKVMN) \cite{zhang2017dynamic}, which may further improve the predictive accuracy. 

\section{Future Directions}
\subsection{LLMs for Knowledge Tracing in Learner Model}

The potential of LLMs for knowledge tracing relies on at least two key aspects: firstly, their capability in identifying knowledge components \cite{koedinger2012knowledge,pavlik2021automatic}, which encapsulate the prerequisite knowledge for proficiently addressing specific questions with the given context; and secondly, their integration with machine learning models (self-selected by LLMs or external), which is further bolstered by LLMs' inherent interpretability, facilitating cohesive reasoning, assessment, and predictive capabilities concerning the learner performance. Further research in this direction holds significant promise for advancing our understanding and application of knowledge tracing methodologies based on LLMs.

\subsection{LLM-based Trace of Learners' Learning for Intelligent Tutoring Systems} 

The present study's finding motivate future research into LLMs to augment and complement modeling of learner learning and dynamic learning states within ITSs. The endeavor involves utilizing diverse data types, including numerical, textual, and even multimodal inputs, to construct a comprehensive learner model. Drawing from LLMs like ChatGPT, Llama, and Gemini, along with various machine learning methods, future research could provide effective real-time prediction of learner learning. By integrating insights from LLMs and machine learning, this approach enhances the pedagogical component of Intelligent Tutoring Systems, enabling more precise instructional strategies and feedback mechanisms. Specifically, the present use of LLMs for learner modeling could be used in nascent applications for tutoring through Expectation-Misconception Tailored (EMT) conversation styles in adult literacy \cite{graesser2004autotutor,graesser2005computerized,ahmed2023chatgpt} or applications of LLMs to automatically generate peer-tutoring dialog\cite{schmucker2023ruffle}. Improving learner modeling through these applications could enable more personalized and effective pedagogical strategies and feedback for learners. 

\section{Conclusion}

The present study investigates the use of LLMs, specifically GPT-4, in predicting learner's learning performance in the context of adult literacy in Intelligent Tutoring Systems. We developed an LLM-based prediction method that integrates the encoding of numerical learning performance data into a contextual prompt, conducting data analysis and prediction within LLM component, and decoding the output to obtain predicted learning performance data. Compared to traditional machine learning methods, such as BKT, PFA, SPARFA-Lite and Tensor Factorization, our LLM-based method achieves higher predictive accuracy when incorporating with XGBoost (selected by GPT-4 in our study).  Although XGBoost running on a local machine initially surpasses GPT-4 in accuracy, the optimization of XGBoost parameters selected by GPT-4 and fine-tuned within the LLM environment exhibits enhanced performance over local execution. Our examination of hyperparameter tuning by GPT-4 versus manual grid search reveals similar outcomes, yet the GPT-4 method introduces a degree of variability. Our findings underscore the potential of merging LLMs with established machine learning frameworks to boost personalization and efficacy in adult literacy education. This work lays the groundwork for future inquiries into the integration of LLMs in ITS environments, demonstrating the increased practicality of employing LLMs for learner performance prediction in AI-enhanced educational contexts.

\begin{credits}
\subsubsection{\ackname} We extend our heartfelt gratitude to Prof. Kenneth R. Koedinger of Carnegie Mellon University, along with Prof. Arthur C. Graesser and Prof. Philip I. Pavlik Jr. from The University of Memphis, for their invaluable insights that significantly enriched this study. Our study was supported by the Institute of Education Sciences, U.S. Department of Education, through Grant R305A200413 and R305A190522 to The University of Memphis. The opinions expressed are those of the authors and do not represent views of the Institute or the U.S. Department of Education. 
\end{credits}

%
%
%
%

\printbibliography

\clearpage
\appendix
\section*{Appendix A: Example Prompt Strategy for Predicting Learning Performance}
\label{appendix:A}

\scriptsize
\begin{longtable}{>{\raggedright}p{6cm}|>{\raggedright\arraybackslash}p{6cm}} 
    \hline
    \textbf{User} & \textbf{GPT-4} \\
    \hline
    \endfirsthead 
    \hline
    \textbf{User} & \textbf{GPT-4} \\
    \hline
    \endhead 
    User: Hello ChatGPT, I am working on a project related to adult literacy and have a dataset from the Center for the Study of Adult Literacy (CSAL) AutoTutor lessons. This dataset contains responses from adult learners engaged in reading comprehension activities. I seek your expertise in data mining and analysis to uncover insights and predict learning outcomes. & I'd be delighted to help you with your project on adult literacy. To get started and provide you with the most accurate assistance, I'll need some additional information about your dataset and your specific goals. Here are a few questions to guide our next steps: 
    \begin{itemize}
        \item Dataset Overview: Could you provide an overview of the dataset? For instance, what are the features (columns) in your dataset, and what types of data do they contain (numerical, categorical, text, etc.)?

        \item Learning Performance: What specific learning performance are you interested in predicting? For example, are you looking to predict overall improvement in literacy levels, comprehension scores after lessons, or something else?

        \item Analysis Goals: Beyond predicting learning performance, are there any specific insights or patterns you're hoping to uncover? This might include identifying the most challenging lessons, understanding which factors contribute most to literacy improvement, or anything else relevant to your project.
    \end{itemize}

    Once you provide more details, we can discuss potential data mining techniques and analytical approaches that could be beneficial for your project ...
    \\
    \hline
    User: Here is the information for one CSAL lesson. The topic of the lesson resolves around Problems and Solution for testing adult reading comprehension. The reading content is titled 'Minor Burns'. We assume the difficulty level of this lesson and questions as Medium.
    
    The reading material for the current lesson is in following:\colorbox{yellow}{\{Reading Material\}}. All the questions and corresponding answers are as follows: \colorbox{yellow}{\{Questions\}}. Could you provide a mapping of the relationship between the knowledge required for each question, based on the lesson content, and the reading skills necessary to comprehend and master this knowledge? & GPT-4: To map the relationship between the knowledge required for each question based on the "Minor Burns" lesson content, we'll examine the questions and identify the specific pieces of information or skills needed to answer them correctly. This process involves linking each question to the relevant parts of the text and categorizing the type of knowledge or cognitive skill they test, such as factual knowledge, comprehension, application, or analysis. Knowledge Mapping: 
    \begin{itemize}
        \item Question 1: What is the topic of this text? \newline Knowledge Required: Factual knowledge about the text's main topic. \newline 
        Related Text Content: The entire reading is focused on minor burns. \newline
        Skill Type: Identification of main idea. 
    \end{itemize} 
    \(\vdots\) \newline
    This mapping highlights that successfully answering the questions requires a mix of factual knowledge, comprehension, application, analysis, and evaluation skills. Each question is tied to specific content within the "Minor Burns" lesson, demanding learners to engage with the material at different cognitive levels ...
    \\ 
    \hline
    User: This dataset is organized into five distinct sections, each comprising a training and a testing dataset. These datasets feature four key columns: learner ID, Question ID, Attempts, and Obs (actual performance). Within the Obs column, a performance score of 0 signifies a completely incorrect answer, while a score of 1 denotes a completely correct answer, in relation to the previously mentioned questions and their respective answers. All historical performance data are encapsulated within the training dataset, whereas data awaiting prediction are contained within the testing dataset. The task involves utilizing the training dataset to fine-tune the GPT-4 model, applying the testing dataset for validation, and predicting learner performance as reflected in the Obs column. \newline 
    Notes for consideration: 
    \begin{enumerate}
        \item I should rely on my analytical and computational skills for making these predictions. 
        \item It's requested that I calculate a continuous likelihood score between 0 and 1 for each learner's chance of correctly answering a question, based on the learner ID and Question ID. This score should directly correlate with the likelihood of a correct or incorrect answer, adhering to the criteria outlined in the dataset's description. 
    \end{enumerate} 
    \{Add ``Avoid using the machine learning models." for GPT-4 model.\}
    \colorbox{yellow}{\{Attached Dataset Files\}}
    & GPT-4: To begin analyzing and predicting learner performance using the provided datasets, I'll start by examining one of the training datasets to understand its structure and content. This initial analysis will help me determine the appropriate data preprocessing steps and model training approach... \newline 
    \colorbox{yellow}{\{Compile Dataset Files\}}
    \newline
    \vdots \newline 
    Given the nature of the data (categorical IDs for learners and questions without additional metadata), the model will need to capture patterns based on historical performance metrics associated with specific learners and questions. 
    \newline
    For predicting learner performance on reading comprehension questions based on the dataset provided, we can explore a variety of methods. These can range from heuristic approaches to sophisticated machine learning models.  \colorbox{yellow}{\{Suggest Methods\}} \newline
    Let's outline potential methods within these categories:
    \begin{enumerate}
        \item Heuristic Methods ...
        \item Machine Learning Models ...
    \end{enumerate} 
    \vdots
    For simplicity and efficiency, we'll begin with a model that doesn't require extensive preprocessing of these categorical features. A gradient boosting machine, such as XGBoost, can handle numerical categorical variables directly and is well-suited for this task due to its ability to model complex relationships and provide probability scores for outcomes. \newline
    \ \ \ \ \vdots
    \newline
    \{Alternatively, GPT-4 is capable of autonomously selecting the optimal prediction method, with logistic regression and XGBoost being the most commonly utilized approaches based on our experiments.\}
\\ \hline
This is one optional step:
\{User: Request to verify dataset information and details to ensure GPT-4 comprehends the data and can perform fundamental statistical analyses. For instance, calculate the number of students, the number of questions, and attempts, including the distribution and frequency of these elements, among others. \} & \{GPT-4: Showcase its understanding and analytical findings in response to your request ...\}
\\ 
    \hline
    User: Could you assist in employing XGBoost regression for model development and prediction tasks across the five folds? I would appreciate if you could also provide the validation outcomes, specifically Root Mean Squared Error (RMSE), for each fold. Additionally, could you detail the specific configuration of the XGBoost algorithm utilized in this analysis?  
    \newline
    \{If you opt to not use machine learning models and instead rely on GPT-4's intrinsic reasoning and computational abilities for prediction, you can request GPT-4 to proceed using its heuristic approach in formulating responses.\} & GPT 4: Using a very simplified XGBoost regression model for the first fold, we successfully completed the training and evaluation process within the computational constraints. Here are the updated results for all five folds in following: 
    \begin{itemize}
        \item Fold 1:
        \item Fold 2:
        \item Fold 3:
        \item Fold 4: 
        \item Fold 5: 
    \end{itemize}
    \vdots
    \\ \hline
    User: Can you help me present the model
setting? & GPT-4: For the evaluations performed across the five folds, the XGBoost regression model was configured with the following settings to balance the execution time and predictive performance within the operational constraints: 
\begin{itemize}
    \item n\_estimators:... 
    \item max\_depth:...
    \item learning\_rate:
\end{itemize}
\vdots
\\ \hline
User: Is there any reading comprehension skills for answering the questions? Can you help me assess the learners reading comprehension skills? & GPT-4: To assess a learner's reading comprehension skills based on the performance data provided, I will first need to analyze the dataset for patterns in their responses to questions that we can infer are related to reading comprehension. Given the structure of the dataset, I can attempt to identify: ... \\ \hline
\colorbox{yellow}{\{Fine-tuning Model\}} \newline
User: Could you assist in optimizing the model by fine-tuning its hyperparameters to achieve enhanced predictive performance? & GPT-4: Optimizing an XGBoost model involves tuning various hyperparameters to find the best combination that minimizes prediction error on your validation set. The most common hyperparameters to adjust include: ... \newline 
Let's start by selecting a set of hyperparameters to tune for the XGBoost model...
\\ \hline
User: \{Other prompts that you need\} & GPT-4: \{Other responses\}
\\
    \hline
    
    \label{tab:iterative-conversation}
\end{longtable}

\end{document}